\documentstyle[12pt,epsfig]{article}
\newcommand{\Frac}[2]%
{{\textstyle \frac{\mbox{\footnotesize $#1$}\rule[-0.9mm]{0mm}{1mm}}%
{\mbox{\footnotesize $#2$}\rule{0mm}{3.1mm}}}}
\renewcommand{\thefootnote}{\fnsymbol{footnote}}

\begin{document}
\begin{titlepage}
\vspace*{-12 mm}
\noindent
\begin{flushright}
\begin{tabular}{l@{}}
\end{tabular}
\end{flushright}
\vskip 12 mm
\begin{center}
{\large \bf 
Neutrinos from spin dynamics
}
\\[14 mm]
{\bf Steven D. Bass}
\footnote[1]{steven.bass@cern.ch}
\\[10mm]   
{\em Particle Theory Unit, Physics Department, 
     CERN, CH 1211 Gen{\`e}ve 23, Switzerland}
\\[5mm]
{\em 
Institute for Theoretical Physics, \\
Universit\"at Innsbruck,
Technikerstrasse 25, Innsbruck, A 6020 Austria
\\[5mm]
}
\end{center}
\vskip 10 mm
\begin{abstract}
\noindent
We conjecture that neutrino physics 
might correspond to the spontaneous magnetisation phase of an 
Ising-like spin model 
interaction coupled to neutrino chirality which operates at scales 
close to the Planck mass.
We argue that 
this scenario
leads to a simple extension of the Standard Model 
with no additional parameters that
dynamically generates parity violation and 
spontaneous symmetry breaking for the gauge bosons which couple to the 
neutrino.
The neutrino mass in the model is 
$m_{\nu} \sim \Lambda_{\rm ew}^2 / 2M$ 
where $\Lambda_{\rm ew}$ is the electroweak scale and $M$ 
is the scale of the ``spin-spin'' interaction.
For the ground state of the model
the free energy density corresponding to the cosmological constant is 
$\rho_{\Lambda}^{1 \over 4} \sim m_{\nu}$,
consistent with observation.

\end{abstract}
\end{titlepage}
\renewcommand{\labelenumi}{(\alph{enumi})}
\renewcommand{\labelenumii}{(\roman{enumii})}
\renewcommand{\thefootnote}{\arabic{footnote}}
\newpage
\baselineskip=6truemm

\section{Introduction}

The neutrino sector is one of the most puzzling in particle physics.
Only left-handed neutrinos participate in weak interactions.
Recent experiments have revealed oscillations between the three 
families of neutrino 
plus small neutrino masses, 
with the heaviest neutrino mass $\sim 0.05$ eV, 
much less than the masses of the charged leptons and quarks.
The mass of the lightest neutrino is presently not well constrained
although for a normal hierarchy of 
neutrino masses 
with
$m_1 \ll m_2 \ll m_3$ 
one finds $m_1 \ll m_2 \sim 0.008$ eV
-- for recent reviews see 
Refs.\cite{conrad,altarelli,whitepaper,kayser,blondel}.
Possible explanations for the small neutrino mass involve either 
right-handed sterile neutrinos 
(together with some additional ``new physics'' to suppress the mass 
 relative to the charged leptons)
or 
Majorana mass terms with local coupling to scalar Higgs fields
using the see-saw mechanism \cite{seesaw}
to push the mass of the right-handed neutrinos to a very high scale,
thus connecting the neutrino sector with new physics at much higher
mass scales.

It is interesting to ask whether there might be an alternative 
explanation:
Can we construct a dynamical mechanism which yields observed 
neutrino physics ? 
Is there an analogue situation in other branches of physics that 
one might hope to learn from ?

Here we investigate a possible analogy with the Ising model of statistical 
mechanics and, more generally, spin glass.
Suppose that we associate the ``spins'' in the Ising model with neutrino 
chirality and the ``internal energy per spin'' with the neutrino mass.
The ground state of the Ising model exhibits spontaneous magnetisation 
where all the ``spins'' line up;
the ``internal energy per spin'' and the free energy density of the spin 
system go to zero.

This observation suggests the possibility that, perhaps, neutrino physics 
might arise from collective spin model phenomena 
at a large scale, 
logarithmically close to the Planck mass.
Electroweak interactions 
might be included as an ``impurity'' in the model: 
the $\nu \rightarrow e W$ process corresponds to a small 
but finite probability for one of the ``spins'' to turn off if 
the Ising interaction couples just to neutrinos.
The spontaneous magnetisation phase 
then exhibits parity violation
as the right-handed neutrino decouples from the physics.
We argue that it also exhibits spontaneous breaking of 
the SU(2) gauge symmetry coupled to the neutrino.
The neutrino mass $m_{\nu}$
in this picture is expected to be about $\Lambda_{\rm ew}^2/2M$
where 
$\Lambda_{\rm ew}$ is the electroweak scale and $M$ is the scale of 
the spin model interaction.
The free energy density of the spin system behaves like a cosmological 
constant and for the low energy spontaneous magnetisation 
phase is
$\rho_{\Lambda}^{1 \over 4} \sim m_{\nu} +O(kT)$, 
where $kT \sim 0.0002$ eV for the CMB temperature of free space.

\section{Ising model dynamics}

We first briefly outline the basics of the Ising model 
(for reviews see \cite{creutz,huang})
and then discuss similarities and possible application to neutrino physics.
The Ising model uses a spin lattice to 
study ferromagnetism for a spin system in thermal equilibrium.
Applications include crystals, lattice gases and spin glass.
One assigns a ``spin'' ($= \pm 1$) to each site and 
introduces a nearest neighbour ``spin-spin'' interaction.
In two dimensions the Hamiltonian reads
\begin{equation}
H = 
- J \ \sum_{i,j} \ 
( \sigma_{i,j} \sigma_{i+1,j}  +  \sigma_{i,j+1} \sigma_{i,j} )
\ 
- 
\ h \sum_{i,j} \sigma_{i,j}
.
\end{equation}
Here $J$ is the bond energy and
$h= \mu B$ where $B$ denotes any external magnetic field and $\mu$
is the magnetic moment.
(In this paper we take $h=B=0$.)
One sums over the possible spins $\sigma_{ij}$.
Physical observables are calculated through the partition function 
\begin{equation}
Z = \sum_{\sigma_{ij}} \exp ( - \beta H )
.
\end{equation}
Here
$
\beta = 1 / k T
$
where $k$ is Boltzmann's constant and $T$ is the temperature;
$
k = 1.38 \times 10^{-23} \ J K^{-1} = 8.617 \times 10^{-5} \ eV \ K^{-1}
$
and 
$
k T|_{300 K} 
= 
[ 38.68 ]^{-1} eV
$.
We can normalise the energy by adding a constant so that neighbouring 
parallel spins give zero contribution.
Then, the only positive contribution to the energy will 
be from neighbouring disjoint spins of $2J$ 
and the probability for that will be $\exp (- 2 \beta J)$.
Once a magnetisation direction is selected, 
it remains stable because of the infinite number of degrees of freedom 
in the thermodynamic limit.

The ``internal energy per spin'' corresponding to the Hamiltonian in 
Eq.(1) is
\begin{equation}
\epsilon (\beta J) 
=
- 2J \tanh (2 \beta J)
+ {K \over \pi} {d K \over d \beta} 
\int_0^{\pi \over 2} d \phi {\sin^2 \phi \over \Delta (1 + \Delta) }
\end{equation}
where
\begin{equation}
K = {2 \over \cosh (2 \beta J) \coth (2 \beta J) }
\end{equation}
and
\begin{equation}
\Delta = \sqrt{ 1 - K^2 \sin^2 \phi }.
\end{equation}
The ``free energy per spin'' or free energy density for the system is
\begin{equation}
F (\beta) = 
- {1 \over \beta} \biggl[
\ln ( 2 \cosh 2 \beta J )
+ {1 \over 2 \pi}
\int_0^{{\pi \over 2}} 
d \phi \ln {1 \over 2} ( 1 + \sqrt{1 - K^2 \sin^2 \phi} )
\biggr] 
.
\end{equation}
The internal energy and the free energy are related 
through
\begin{eqnarray}
\epsilon &=& {\partial \over \partial \beta} \ ( \ \beta F \ )
\nonumber \\
\epsilon &=& F + TS
\end{eqnarray}
where $S$ is the entropy.
The pressure for the spin system is $p = -F$.
The Ising model has a second order phase transition and 
exhibits 
spontaneous magnetisation.
In two dimensions the critical coupling
$(\beta J)_c$
is determined through the equation
$
\{\sinh 2 (\beta J)_c = 1 \}
$.
For values of  $(\beta J) \geq (\beta J)_c$
the magnetisation per spin is $\pm {\cal M}$
where
\begin{equation}
{\cal M} 
=
\biggl\{ 
1 - [ \sinh ( 2 \beta J ) ]^{-4} \biggr\}^{1 \over 8} 
.
\end{equation}
The magnetisation vanishes for $(\beta J) < (\beta J)_c$.
In the limit $\beta J \rightarrow \infty$, 
one finds
\begin{equation}
\epsilon (\beta J) = - 2 J - 24 J \exp{(-8 \beta J)} + ...
\end{equation}
\begin{equation}
F =  -2 J + 3 kT \exp{(-8 \beta J)} + ... 
\end{equation}
and
\begin{equation}
{\cal M} = 1 - 2 \exp{(-8 \beta J)} + ...
\end{equation}
As advertised above,
the zero-point energy is then renormalised by adding $+2 J$ 
to 
$\epsilon (\beta J)$ and $F$ 
(or $2J N$ to the Hamiltonian where $N$ is the number of spin sites)
so that the
``internal energy per spin'' and free energy density vanish in the 
ground state with spontaneous magnetisation: $\epsilon (\infty) = 0$.
In general, the critical coupling depends on the number of dimensions.
For 1, 2, 3 and 4 
Euclidean dimensions the critical coupling 
$(\beta J)_c$
is
$\infty$, 0.441, 0.167 and 0.150 respectively.

One can extend the model to spin glasses \cite{anderson}
by introducing a probability distribution over the parameters of the model,
e.g. the bond energies $J$ and the magnetic moments.

\section{Neutrinos}

How can we construct an analogy with neutrinos and particle physics ?

First, the Ising-like interaction itself must be non-gauged otherwise 
it will average to zero and 
there will be no spontaneous symmetry breaking and no spontaneous
magnetisation 
(see e.g. page 51 of \cite{creutz}).

Second, 
it is necessary to set a mass scale for $J$.
If the Ising model analogy is to have connection with particle physics 
it is important to note that the coupling constant for the ``spin-spin'' 
interaction is proportional to the mass scale $J$.
It therefore cannot correspond to a renormalisable interaction 
suggesting that fluctuations around the scale $J$ 
occur only near the extreme high-energy or 
high-temperature limit of particle physics near the Planck mass $M$.
We consider the effect of taking $J \sim + M$.
The combination $\beta J$ is then very large making it almost 
certain that, 
if the analogy is applicable, the spontaneous 
magnetisation phase is 
the one relevant to particle physics phenomena.
(Fermion generations in particle physics might be a further hint
 at some kind of spin related dynamics at a very high mass scale.) 
The exponential suppression factor $e^{-2 \beta J}$ ensures 
that 
fluctuations associated 
with the Ising-like interaction are negligible, 
thus preserving renormalisability for all practical purposes.

Motivated by these observations, suppose we start with a gauge theory 
based on 
\begin{equation}
SU(3) \otimes SU(2) \otimes U(1)
\end{equation}
coupled to quarks and leptons with no chiral dependent couplings and
unbroken local gauge invariance.
We 
then turn on the non-gauged
Ising interaction coupled just to the neutrino in the upper 
component of the SU(2) isodoublet with the 
coupling $J \sim M \gg \alpha_s, \alpha_{\rm ew}, \alpha$
(the QCD, SU(2) weak and QED couplings).
It seems reasonable that the Ising interaction here exhibits 
the same two-phase picture with spontaneous magnetisation.
Then, in the symmetric phase where $\beta J < (\beta J)_c$ 
the theory is symmetric under exchange of left and right 
handed neutrino chiralities and we have unbroken local gauge invariance.
In the spontaneous magnetisation phase the neutrino vacuum is 
``spin''-polarised,
a choice of 
chirality is made and 
the right-handed neutrino decouples from the physics.
Parity is spontaneously broken
and 
the gauge theory coupled to the leptons becomes
\begin{equation}
SU(2)_L \otimes U(1)
\end{equation}
It is reasonable to believe that the SU(2) 
gauge symmetry coupled
to the neutrino is now spontaneously broken.
To understand how the W$^{\pm}$ and Z$^0$ 
gauge bosons 
might acquire mass,
first consider the issue of confinement. 
Before we turn on the Ising interaction we have QCD quark and 
gluon 
confinement plus a 
vector SU(2) gauge theory with electron and neutrino confinement.
Confinement is intimately connected with dynamical chiral symmetry 
breaking 
and the generation of Dirac mass terms for 
constituent quarks and (at this stage) analogue constituent leptons.
(See Ref. \cite{alkofer} for a recent discussion in QCD 
\footnote{ 
Pure Yang-Mills theory and Yang-Mills theory coupled to fermions
are both confining theories but the mechanism is different for each.
Recent calculations \cite{alkofer} of quenched QCD in the covariant
Landau gauge 
suggest that if chiral symmetry is restored, 
then the quark confinement solution disappears.
}
and 
 Ref. \cite{bag} 
 for an overview how this connection 
 is implemented in the phenomenological Bag model of confinement.)
Next turn on the Ising interaction and go to the spontaneous 
magnetisation phase.
The right-handed neutrino decouples from the physics:
the scalar chiral condensate 
for the neutrino 
``melts'' and the confining solution for the neutrino should disappear.
The left-handed neutrino will want to escape the confinement radius 
whereas the SU(2) gauge bosons 
will want to remain confined 
-- in contradiction to the fundamental SU(2) symmetry.
Some modification of the propagators must happen,
{\it viz.} 
mass generation for the gauge bosons and 
the transition from the confinement to Higgs phases of the model
so that the 
Coulomb force is replaced by a force of finite range 
with finite mass scale 
and the issues associated with infrared slavery are avoided
\footnote{A similar effect occurs in 1+1 dimensional gauge theories 
such as the Schwinger model coupled to dynamical fermions: 
confinement gives way 
to Higgs phenomena if the fermion mass is set exactly to zero \cite{gross}.}
.
In this scenario 
experimental investigations of electroweak symmetry breaking will 
be probing fundamental properties of strong coupling dynamics.
The W$^{\pm}$ and Z$^0$ gauge bosons which couple 
to the neutrino 
are massive and the QED photon and QCD gluons are massless.

What about the electroweak scale ? 
Is there anything which stabilises it ?
First,
in this scenario 
the SU(2) gauge symmetry coupled to the neutrino is 
dynamically broken.
Second,
dynamics 
associated with the Ising ``spin'' interaction are 
suppressed
in the spontaneous magnetisation phase by the exponential factor
$e^{-\beta M}$ where
$\beta$ is a finite inverse mass-scale
like an inverse small-temperature factor
or a mass scale typical of laboratory experiments,
suggesting a natural hierarchy of scales.
Further, the enormous excitation energy for right-handed neutrinos
$\sim M$ reminds us of the infinite energy required to excite a
free-quark in QCD because of confinement.
The infinite free-quark excitation energy compares
with the finite chiral symmetry breaking scale $F_{\pi} \sim 100$ MeV.
Likewise, the electroweak mass-scale
$\sim 250$ GeV is much less than the right-handed neutrino excitation energy.

What about finite neutrino masses ?
Weak interactions mean that we have two basic scales in the problem: 
$J \sim M$ and 
the electroweak scale $\Lambda_{\rm ew}$ 
induced by spontaneous symmetry breaking.
(Realistic accessible neutrino kinetic energies will
 be much less
 than $J \sim M$ and hence a minor correction to the total
 energy and Hamiltonian.)
First, make the usual assumption that electroweak symmetry breaking
generates Dirac mass terms 
for fermions which participate in electroweak interactions,
e.g. as ``impurities'' in the ``spin'' system.
Next, suppose that we 
{\it approximate} 
the two-phase spin-magnetisation system 
by left-handed and right-handed neutrinos with Majorana mass terms
so that the right-handed neutrino appears only at scales 
$\sim O(M)$ and 
electroweak processes contribute a regular Dirac mass term.
Then the ``internal energies per spin'' read in matrix form as
\begin{equation} 
M_{\rm \nu}
\sim 
\left[ \begin{array}{cc}
\! 0  &   \Lambda_{\rm ew}  \! \\
\! \Lambda_{\rm ew} &  2M  \!
\end{array} \right] 
\end{equation}
where the first row and first column refer to the left-handed 
states of the neutrino and the second row and second column 
refer to the right-handed states.
Diagonalising this matrix for $M \gg \Lambda_{\rm ew}$ gives
the light mass eigenvalue
\begin{equation}
m_{\nu} \sim \Lambda^2_{\rm ew} / 2M
\end{equation}
after the usual chiral rotation.
Substituting the values of the Planck mass
$M_{\rm Pl} \equiv \sqrt{\hbar c / G_N} \sim 1.2 \times 10^{19}$ GeV 
and the electroweak scale
$\Lambda_{\rm ew} \sim 250$ GeV
into Eq.(15)
gives $m_{\nu} \sim 3 \times 10^{-6}$ eV,
which is plausible for the mass of 
the lightest neutrino and respectable given the simple approximations 
used above.

The matrix in Eq.(14) looks like the see-saw mechanism result
\cite{seesaw} although the fundamental physics is quite different.
In the see-saw picture the left-handed and right-handed components
of a four-state 
Dirac neutrino are split by Majorana mass terms involving coupling 
to scalar Higgs fields
into a pair of
two-state Majorana neutrinos with different masses.
The masses of these 
left-handed light and 
right-handed heavy Majorana neutrinos are related through
$
m_{\nu} 
\sim \Lambda^2_{\rm ew} / M_D 
$
where $m_{\nu}$ is the mass of the light neutrino and 
$M_D$ is both the ``new physics'' scale and the mass of the heavy neutrino.
For a light neutrino mass $\sim 0.05$ eV this relation gives 
$M_D \sim 10^{15}$GeV,
which is not 
so far (on a logarithmic scale) 
from the Planck mass $M_{\rm Pl} = 1.2 \times 10^{19}$ GeV.
In the two-phase picture suggested here the tiny mass for 
the neutrino originates from collective ``spin dynamics'' 
near the Planck scale instead of through local 
Yukawa couplings to elementary scalar Higgs fields. 
The connection to the see-saw matrix (14) suggests that, perhaps,
the 
neutrino in this picture should be Majorana and
therefore have no vector current: its electric charge should vanish.

It is interesting that the vacuum energy density corresponding 
to the 
cosmological constant 
$\rho_{\Lambda}^{1 \over 4} \sim 0.002$ eV 
\cite{lambda}
has a similar numerical value to the range of 
possible light neutrino masses \cite{altarelli}, 
prompting the question whether related underlying dynamics 
might be at work ?
In the Ising model the free energy density for the spin system 
(or energy available for work)
is related 
to the internal energy per spin through Eq.(7):
$\epsilon = {\partial (\beta F) / \partial \beta}$.
It follows that 
$F \sim \epsilon \sim {\Lambda_{\rm ew}^2 / 2M} + O(kT)$ 
where 
$T$ is the temperature of the system,
which for the 
present Universe is the CMB temperature 2.73K or $kT \sim 0.0002$ eV.
The free energy density is suppressed by the ``spin dynamics'' 
which generate the spontaneous magnetisation and which 
``spin''-polarise the neutrino vacuum.
The small finite value reflects the relatively small scalar component, 
$\sim \Lambda_{\rm ew} / 2 M 
 \ 
 | {\rm scalar} \rangle_{\rm ew, qcd}$,
associated with the electroweak and QCD scales which is 
induced in an otherwise ``spin''-polarised total vacuum,
{\it viz.}
\begin{equation}
| \ {\rm vacuum} \ \rangle_{\rm total} 
\sim 
(\Lambda_{\rm ew}/2M) \
| \ {\rm scalar} \ \rangle_{\rm ew, qcd}
\
+
\ 
| \ {\rm polarised } \ \rangle_{\nu}
\end{equation}
 Electroweak and QCD interactions couple to the scalar condensates 
 associated with the scales
 $\Lambda_{\rm ew}$ and $\Lambda_{\rm qcd}$
 whereas the total vacuum in this picture is dominated 
 by the ``spin''-polarised neutrino component with zero free-energy density.
 Without the ``spin-polarised'' component 
 the free-energy density would be
 $\rho_{\Lambda}^{1 \over 4} \sim \Lambda_{\rm ew}$.
The positive sign for the free-energy density corresponds 
to negative pressure.

In conclusion, there are clear similarities between neutrino 
phenomenology and spontaneous magnetization in the Ising model.
This suggests the conjecture that, perhaps, neutrinos are associated 
with the spontaneous magnetisation phase of a spin model 
interaction which operates at scales close to the Planck mass. 
The phase transition associated with the spontaneous magnetisation of 
neutrino chirality in the spin system would, 
in turn, lead to parity violation and,
we have argued,
also
spontaneous breaking of 
the SU(2) gauge symmetry coupled to the neutrino.
Further, there would be a rapid drop in the potential governing 
the
effective neutrino mass or 
``energy per spin'' 
from a value close to $M$ at temperatures $kT \sim M$ 
to a value $\sim \Lambda_{\rm ew}^2 / 2M$ 
in the ground state which might be connected to the potential needed 
for inflation.
There is no elementary scalar field in the model.
It is interesting that the observed cosmological constant 
corresponds to a vacuum energy density 
$\rho_{\Lambda}^{1 \over 4}$
comparable with the expected value of the light neutrino mass.
For the spin system 
the corresponding free energy density goes as $F \sim m_{\nu} + O(kT)$.

\section*{Acknowledgements}

I thank the Austrian Science Fund for financial support (grant P17778-N08),
the CERN particle theory unit for hospitality and
T. Ericson, B. Kayser and G. Ross for helpful conversations.

\end{document}